\documentclass[prb,twocolumn,floatfix,amsfonts]{revtex4-1}
\usepackage{graphicx,graphics,color,epsfig}
\usepackage{bm}
\usepackage{amsmath}
\usepackage{amssymb}
\usepackage{color}

\newcommand{\YRS}{YbRh$_2$Si$_2$}

\begin{document}
\title{Comment on
``Tuning low-energy scales in ${\bf {\rm YbRh_2Si_2}}$ by non-isoelectronic
substitution and pressure''}

\author{Steffen Wirth}
\affiliation{Max Planck Institute for Chemical Physics of Solids, 01187
Dresden, Germany}

\author{Silke Paschen}
\affiliation{Institute of Solid State Physics, Vienna University of
Technology, 1040 Vienna, Austria and\\
Department of Physics and Astronomy, Rice Center for Quantum
Materials, Rice University, Houston, Texas, 77005, USA}

\author{Qimiao Si}
\affiliation{Department of Physics and Astronomy, Rice Center for Quantum
Materials, Rice University, Houston, Texas, 77005, USA}

\author{Frank Steglich}
\affiliation{Max Planck Institute for Chemical Physics of Solids, 01187
Dresden, Germany and\\
Center for Correlated Matter, Zhejiang University, Hangzhou,
Zhejiang 310058, China and\\
Institute of Physics, Chinese Academy of Sciences, Beijing
100190, China}

\date{\today}

\begin{abstract}
In Ref.\,1, Schubert {\it et al.}\ [Phys.\ Rev.\ Research {\bf 1}, 032004
(2019)] reported measurements of the isothermal magnetoresistance of Fe- and
Ni-substituted \YRS, based on which they raised questions about the Kondo
destruction description for the magnetic field-induced quantum critical point
(QCP) of pristine \YRS. Here we make three points. Firstly, as shown by studies
on pristine \YRS\ in Paschen {\it et al.}\ and Friedemann {\it et al.},
isothermal crossed-field and single-field Hall effect measurements are necessary
to ascertain the evolution of the Fermi surface across this QCP. Because
Schubert {\it et al.}\ did not carry out such measurements, their results on Fe-
and Ni-substituted \YRS\ cannot be used to assess the validity of the Kondo
destruction picture neither for substituted nor for pristine \YRS. Secondly,
when referring to the data of Friedemann {\it et al.}\ on the isothermal
crossover of \YRS, they did not recognize the implications of the crossover
width, quantified by the full width at half maximum (FWHM), being linear in
temperature, with zero offset, over about $1.5$ decades in temperature, from
30\,mK  to 1\,K. Finally, in claiming deviations of Hall crossover FWHM data of
Friedemann {\it et al.}\ from the above linear-in-$T$ dependence they neglected
the error bars of these measurements and discarded some of the data points. The
claims of Schubert {\it et al.}\ are thus not supported by data, neither
previously published nor new (Ref.\,1). As such they cannot invalidate the
evidence that has been reported for Kondo destruction quantum criticality in
\YRS.
\end{abstract}
\maketitle

Quantum criticality is a topic of considerable interest for a variety of
strongly correlated electron systems, with antiferromagnetic heavy fermion
systems representing a prototype. From extensive experimental measurements
across QCPs of several heavy fermion metals, a variety of properties are found
\cite{Schroder,Paschen-nature,
Gegenwart-science,Friedemann-pnas,Kue2003,Cus2003,Tok2009,Kre2009,Fri2009,
Pfau2012,Shishido,Custers,Martelli,Proch,Zhao-NatPhys} to be inconsistent with
spin-density-wave quantum criticality \cite{Hertz,Millis,Moriya}, which is based
on Landau's framework of order-parameter fluctuations. Instead, they support
Kondo destruction quantum criticality \cite{Si,Coleman,Senthil}, which goes
beyond the Landau framework through a critical destruction of the static Kondo
entanglement. In particular, across the magnetic field-induced QCP in
YbRh$_2$Si$_2$, the linear-response Hall coefficient determined from a
crossed-field Hall measurement \cite{Paschen-nature,Friedemann-pnas}, along with
single-field Hall effect \cite{Paschen-nature,Friedemann-pnas},
magnetoresistance \cite{Paschen-nature,Friedemann-pnas}, and thermodynamic
properties \cite{Gegenwart-science}, provided evidence for an extra energy
scale, $T^*$, in the $T$--$B$ plane. This energy scale goes to zero as the QCP
is approached from the non-magnetic side. Isothermal magnetotransport and
thermodynamic properties undergo a rapid crossover across the $T^*$-line, which
extrapolates to a jump in the $T=0$ limit, across generations of YbRh$_2$Si$_2$
samples. These properties are in contrast with the polarization crossover
scenario \cite{Schubert}.

Recently, Schubert {\it et al.}\ \cite{Schubert} studied the magnetoresistance
of Fe- and Ni-substituted YbRh$_2$Si$_2$. Primarily based on the isothermal
behavior of the magnetoresistance in these doped materials, they questioned the
Kondo destruction description for pristine YbRh$_2$Si$_2$. We have the following
comments:

Firstly, the work of Paschen {\it et al.}\ \cite{Paschen-nature}, Gegenwart {\it
et al.}\ \cite{Gegenwart-science}, and Friedemann {\it et al.}\
\cite{Friedemann-pnas}, on pristine YbRh$_2$Si$_2$, is the combination of
systematic studies in terms of magnetoresistance, thermodynamics and, most
notably, single-field and crossed-field Hall measurements. {\em A priori,} only
the latter can directly probe a Fermi surface jump, if also interference from
anomalous Hall contributions can be ruled out. In \YRS, anomalous Hall
contributions were shown to be extremely small \cite{Paschen-nature}. It is also
worth noting that the multiband nature was shown not to be relevant because  a)
the initial isothermal Hall resistivity $\rho_{xy}(B)$ is proportional to the
probing magnetic field [Supporting Information (SI) of Ref.\
\onlinecite{Friedemann-pnas}], implying that one of the multiple bands dominates
the Hall coefficient; and b) on general grounds, a multiband effect {\it per se}
is not relevant to any jump of the Hall coefficient at zero temperature: The
scattering rate by itself has no way of creating a jump -- only the carrier
number can.  An in-depth analysis of the magnetotransport data on the two sides
of the QCP provided a good understanding of the Hall coefficients in terms of
the renormalized bandstructure \cite{SvenPRB}.

Schubert {\it et al.}\ took the unusual approach of assessing the Kondo
destruction physics previously demonstrated for pristine YbRh$_2$Si$_2$
\cite{Paschen-nature,Gegenwart-science,Friedemann-pnas,Proch} by investigating
Fe- and Ni-substituted \YRS. Clearly, it is incumbent upon them to measure these
materials with the same level of rigor previously used for pristine
YbRh$_2$Si$_2$. Most notably, for a new set of samples, the equivalence of
magnetoresistivity measurements and crossed-field Hall effect measurements, as
well as the absence of an appreciable anomalous Hall contribution cannot be
anticipated but must be explicitly demonstrated before drawing any conclusion on
the Fermi-surface evolution across the QCP, which Schubert {\it et al.}\ have
failed to do.

In addition, a more detailed analysis of the effects of disorder introduced by
Fe- and Ni-substitution appears due. The studied substitutions not only
introduce chemical pressure, but also extra carriers and an enhanced degree of
disorder. This leads to rather pronounced changes of the overall
magnetoresistance characteristics [{\it e.g.}, only positive magnetoresistance
for Yb(Rh$_{0.9}$Fe$_{0.1}$)$_2$Si$_2$ and  an only tiny negative contribution
for Yb(Rh$_{0.93}$Fe$_{0.07}$)$_2$Si$_2$], which might indicate  amplified
effects of disorder as well as that another band gets populated by the extra
charge carriers. Also Schubert {\it et al.}'s comparison of the residual
resistivity values cannot add confidence, because the effect of substitutions
cannot be captured by a change in the residual resistivity alone. In the absence
of an understanding of such effects it appears particularly inappropriate to
take crossover fits to such data as evidence against Kondo destruction quantum
criticality not only in their samples, but even in pristine \YRS. Therefore, the
data of Schubert {\it et al.}\ on Fe- and Ni-substituted \YRS, while
interesting  in their own right, can by no means invalidate stringent evidence
for Kondo destruction quantum criticality in pristine \YRS.

\begin{figure}[b]
\centering
\includegraphics[width=0.9\columnwidth]{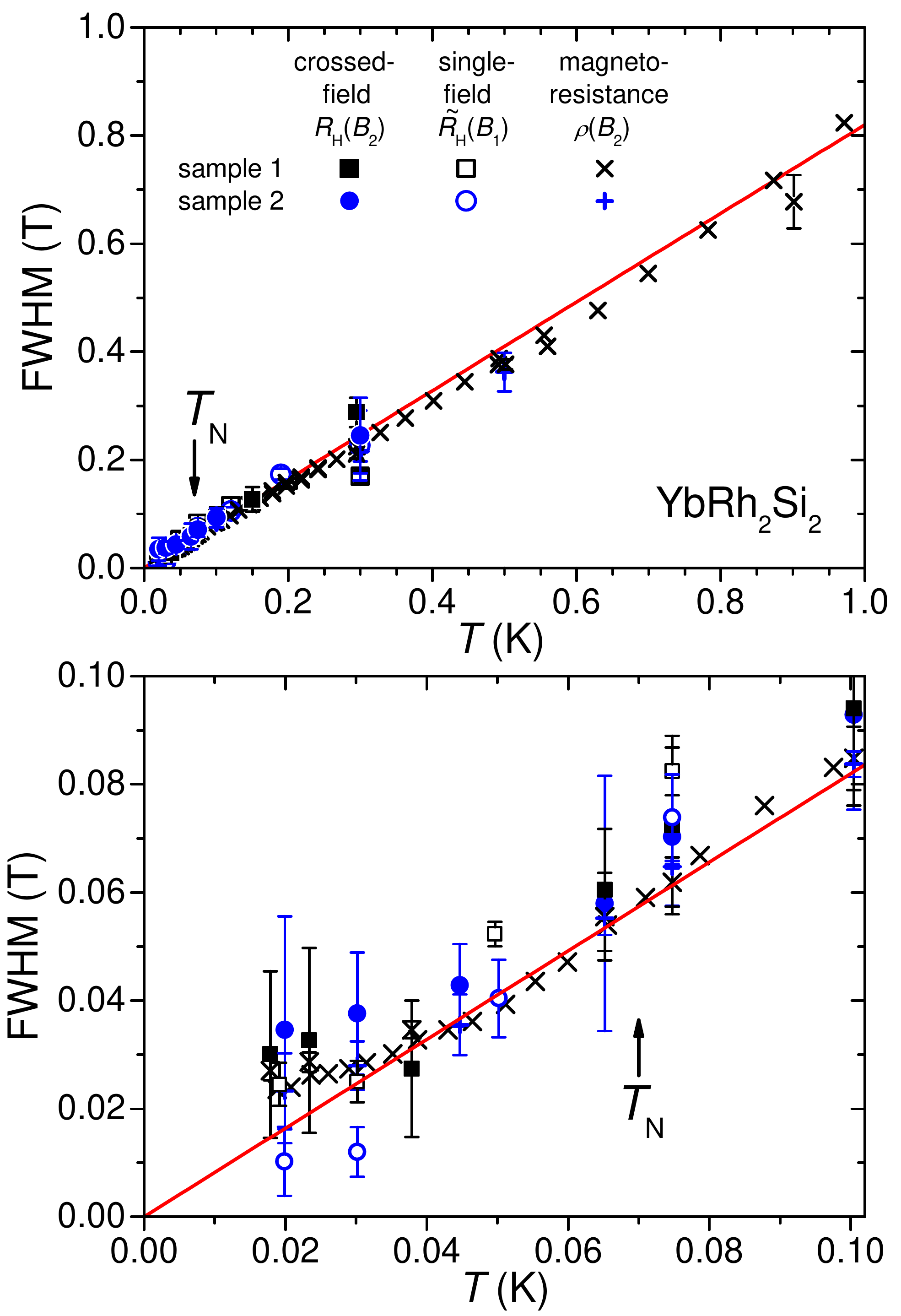}
\caption{\label{Fig1} (a) FWHM {\it vs.}\ $T$ over an extended temperature
range; (b) FWHM {\it vs.}\ $T$ at the lowest measured temperatures. $T_{\rm N}$
indicates the N\'eel temperature. Adapted from Ref.\,\onlinecite{Friedemann-pnas}.}
\end{figure}

To elucidate our second point, we show the stringent $T$-linear FWHM of \YRS\ in
an extended temperature range (Fig.\,\ref{Fig1}a, from Ref.\
\onlinecite{Friedemann-pnas}). Over 1.5 decades in temperature --  from 30\,mK
to 1\,K -- the FWHM is linear in $T$ and extrapolates to zero in the
zero-temperature limit. Along with the finding that for all samples and  all
physical quantities studied the jump size is finite\cite{Friedemann-pnas}, this
makes a clear-cut case that the quantum critical physics within this extended
temperature window is controlled by an underlying QCP for which the Fermi
surface jumps.

To underpin our third point, in Fig.\,\ref{Fig1}b we zoom into the lowest
temperature range of Fig.\,\ref{Fig1}a. Schubert {\it et al.}\ argued that, at
the very lowest measured temperatures ($18\,{\rm mK} \leq T < 30$\,mK), the FWHM
deviates from the linear fit (red line). Yet, within the error bars, this is not
the case. It is clear that extracting the crossover characteristics at these
very low temperatures is complicated by the vicinity to the
antiferromagnetically ordered phase. Classical critical fluctuations associated
with the phase transition will lead to scattering, which will in particular
affect the magnetoresistance. In addition, as pointed out in the SI of
Ref.\,\onlinecite{Friedemann-pnas}, the single-field measurements (including the
blue open circles in Fig.\,\ref{Fig1}b) are obtained with magnetic fields along
the hard magnetic axis (along $c$); magnetic fields in the crossover range are
thus larger than for the other measurements, which facilitates the accurate 
determination of the crossover width (in agreement with the smaller error bars).
Thus, there is no point to selectively discard the single-field Hall effect data
of sample 2 (blue open circles) as Schubert {\it et al.}\ did. Taken together,
there is very solid evidence that the FWHM of the Hall crossover of \YRS\ in the
entire measured temperature range extrapolates to zero in the zero-temperature
limit.

Beyond these main points, it is important to note on the results of
spectroscopic studies of pristine \YRS. STM measurements \cite{Seiro} do detect
a signature at $T^*$ consistent with a critical slowing down (when there is no
$T_{\rm N}$ whatsoever, at $T = 0.3$\,K), contrary to the statement of Schubert
{\it et al.} Equally important, the optical conductivity measured by THz
spectroscopy in pristine \YRS\ has shown $\omega/T$-scaling in the charge
response \cite{Proch}, which is expected in the Kondo destruction picture.

Finally, \YRS\ is not alone in displaying  evidence for  Kondo destruction
quantum criticality. Other examples are the anomalous dynamical scaling observed
over an extended wavevector range in the Brillouin zone of
CeCu$_{5.9}$Au$_{0.1}$ by inelastic neutron scattering measurements
\cite{Schroder}, and evidence for Fermi surface jumps  in CeRnIn$_5$ by dHvA
measurements across its pressure-induced QCP\cite{Shishido}, in
Ce$_3$Pd$_{20}$Si$_6$ based on Hall measurements across its two field-induced
QCPs \cite{Custers,Martelli}, and in CePdAl from Hall measurements across its
line of QCPs in the pressure-magnetic field phase diagram \cite{Zhao-NatPhys}.

\section*{ACKNOWLEDGMENT}
The work was in part supported by the DFG Research Unit 960 `Quantum Phase
Transitions' (S.W. and F.S.), the Austrian Science Fund (FWF project P29296-N27)
and the European Union's Horizon 2020 Research and Innovation Programme, under
Grant Agreement no EMP-824109 (S.P.), and by the NSF grant DMR-1920740 and the
Robert A.\ Welch Foundation grant C-1411 (Q.S.).

\end{document}